%% file: main_text.tex
\documentclass[a4paper, reprint, twocolumn, superscriptaddress]{revtex4-2} %{article}

\usepackage{tikz}
\usepackage[RPvoltages, siunitx]{circuitikz}

\usepackage{graphicx}
\usepackage{subcaption}

\usepackage{amsfonts}
\usepackage{mathtools}
\usepackage{siunitx}

\usepackage[hidelinks]{hyperref}

\usepackage{xcolor}
\hypersetup{
    colorlinks,
    linkcolor={blue!50!black},
    citecolor={blue!50!black},
    urlcolor={blue!80!black}
}

\newcommand{\Uapp}{V}
\newcommand{\Ugap}{U_{\mathrm{g}}}

\newcommand{\Ubr}{U_{\mathrm{b}}}

\newcommand{\Idisch}{j}

\newcommand{\CT}{C_{\mathrm{cell}}}
\newcommand{\Cdie}{C_{\mathrm{d}}}
\newcommand{\Cgap}{C_{\mathrm{g}}}

\newcommand{\dd}{\mathrm{d}}

\begin{document}

\title{Prediction of Re-Ignition Times in Dielectric Barrier Discharges}

\author{C Flores}
\email{cristian.flores@inp-greifswald.de}
\affiliation{Leibniz Institute for Plasma Science and Technology (INP), Greifswald, Germany}
\author{H H\"oft}
\affiliation{Leibniz Institute for Plasma Science and Technology (INP), Greifswald, Germany}
\author{T Hoder}
\affiliation{Leibniz Institute for Plasma Science and Technology (INP), Greifswald, Germany}
\affiliation{Department of Plasma Physics and Technology, Masaryk University, Brno, Czech Republic}
\author{F~X Bronold}
\affiliation{Institute of Physics, University of Greifswald, Greifswald, Germany}
\author{K-D Weltmann}
\affiliation{Leibniz Institute for Plasma Science and Technology (INP), Greifswald, Germany}
\author{M~M Becker}
\affiliation{Leibniz Institute for Plasma Science and Technology (INP), Greifswald, Germany}

\begin{abstract}
Discharge ignition events in dielectric barrier discharges (DBDs) self-organise into spatio-temporal patterns with varying degrees of order.
The complex dynamics of a DBD and intricate structure of occurring patterns complicate the formulation of predictive, mechanistic descriptions.
We present the formulation of a reduced-order model that describes the re-ignition dynamics between consecutive discharges appearing at the same position inside a DBD arrangement.
The model is derived from an equivalent electric circuit and validated against fluid-Poisson simulations and experiments performed with a multi-filament arrangement in air-like gas mixtures at atmospheric pressure driven by sinusoidal high-voltage waveforms.
The experimental scenarios include a highly ordered regime where discharges ignite at regular time and space intervals generating a pattern stable over several periods, and an unstable regime with discharges appearing at seemingly random positions and times.
The model accuracy is assessed in both regimes and it is found that the associated prediction uncertainty provides a quantitative measure of the spatial order of the discharge pattern.
\end{abstract}
\maketitle

\noindent
The order and structure of spatio-temporal ignition patterns in multi-filament DBDs have been investigated extensively for several decades \cite{purwins_dissipative_2010,brunt1991,heitz1999}.
Different order regimes, together with transitions between them, have been observed in both space and time under a wide range of operating conditions and electrode geometries \cite{stollen2006,wild2012,guikema_spontaneous_2000,callegari_pattern_2014, klein_time-resolved_2001}.
It is generally accepted that the behaviour of these systems is governed primarily by the local conditions of the electric field, which is determined by the applied voltage and the distribution of free charges in the gas volume and on the dielectric surface \cite{stollen2006, stollen2007}.
These charges give rise to memory effects that persist between successive discharge events \cite{Hoeft-2014-ID3403,Nemschokmichal-2018-ID5108} and, through their interaction with the applied voltage, contribute to the nonlinear dynamics of the system \cite{brunt1991,kutha2022,li2022}.
Nevertheless, a detailed understanding that enables a predictive and broadly applicable description of this behaviour remains challenging.
This challenge is relevant not only from the perspective of fundamental plasma physics, statistical physics and nonlinear dynamics, but also for practical applications of DBDs.
In applications such as surface treatment or gas conversion (e.g. \cite{cernak2011,douat2023}), the microdischarge density and spatio-temporal distribution often need to be controlled to optimise the performance.

Describing DBDs using electrical equivalent circuit (EC) models is a well-established and continually developing approach.
These models simplify the complex behaviour of the system while retaining its essential predictive capability, in terms of electrical observables \cite{liu_electrical_2003,pipa_simplest_2012}.
Their predictions have been experimentally confirmed for appropriate operating conditions \cite{ivkovic2009,mrkvickova2023}. More advanced EC models can also provide information about the active discharge coverage of the electrode surface area \cite{peeters2015, akishev_memory_2011}; these predictions have likewise been experimentally validated \cite{tyl2021}.
While such information on the spatial characteristics of DBDs is widely used, a comparable approach for describing their temporal behaviour is lacking.

\begin{figure*}[ht]
    %\centering
    \begin{subfigure}{0.4\textwidth}
        \vspace{7mm}
        %\centering
        \input{eec_diagram.tikz}
        \vspace{9mm}
        \caption{}
        \label{fig:eec-diagram}
    \end{subfigure}%
    ~
    \begin{subfigure}{0.6\textwidth}
        %\centering
        \includegraphics[width=\textwidth]{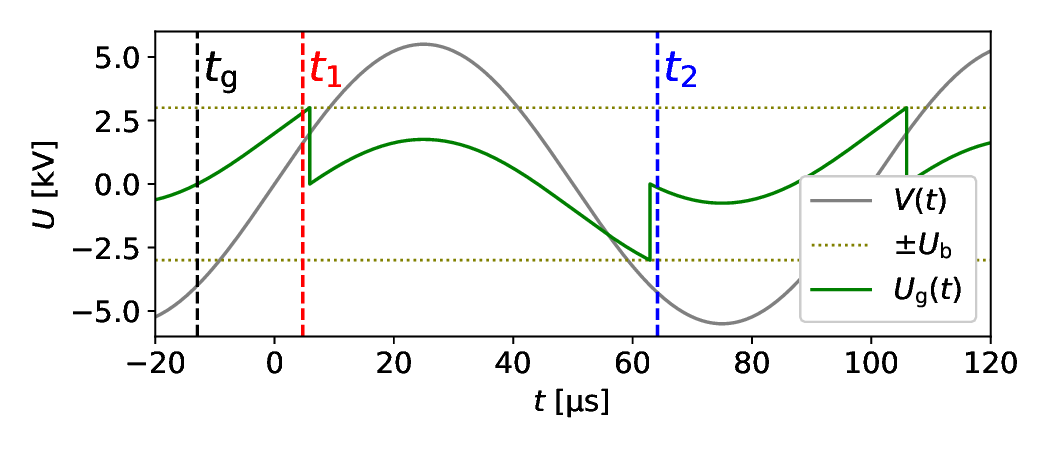}
        \caption{}
        \label{fig:eec-gap-voltage}
    \end{subfigure}
    \caption{(a) Diagram of the EC where the discharge current $\Idisch(t)$ is modeled as a current source. (b) Applied voltage $\Uapp(t)$ and EC results for the voltage across the gap $\Ugap(t)$ during a single period. The vertical line at $t_1$ indicates the instant shortly before the first discharge, while the line at $t_2$ is the instant immediately after the second discharge. The conditions at these instants are used to approximate the discharge ignition times within a single period.}
    \label{fig:equivalent-electric-circuit}
\end{figure*}

In this Letter, we show that an EC model provides a direct relation between the ignition times of discharges in a DBD, and the effective capacitances, the applied-voltage waveform and the voltage across the gap.
Following the work of Willebrand \cite{willebrand_spatio-temporal_1991}, Guikema \cite{guikema_spontaneous_2000} and Klein \cite{klein_time-resolved_2001}, we consider a DBD configuration with an effectively one-dimensional lateral geometry, in which the filaments are constrained to a line perpendicular to the applied electric field and neighbouring filaments can be readily identified \cite{boeuf_generation_2012}.
The EC predictions are validated against temporally and spatially resolved optical measurements and simulations based on a fluid-Poisson (FP) model.

We utilise the simplest formulation of an EC model of a DBD \cite{liu_electrical_2003,pipa_simplest_2012}.
The dielectrics and the gap are modelled as a series of the effective capacitances $\Cdie$ and $\Cgap$, respectively; a third component is connected in parallel to $\Cgap$ to model the discharge current $\Idisch(t)$ (Fig.~\ref{fig:eec-diagram}).
Note, that such formulation can be understood both as an effective global electrical description of multi-filament DBDs (e.g. in ozone generators or large systems for surface treatment, see e.g. \cite{pipa2019}) as well as a 0D detailed description of a single spatially localised discharge ignition \cite{sewraj2011}.
The voltage across the gap $\Ugap(t)$ can be expressed as
\begin{equation}
    \Ugap(t) = \frac{\CT}{\Cgap}\Uapp(t) - \frac{1}{\Cgap + \Cdie}\int^t_{0}\Idisch(\tau)\,\dd \tau + \Ugap(t=0), \label{eq:eec-gap-voltage}
\end{equation}
where $\Uapp(t)$ is the external voltage source driving the DBD and $\CT = (\Cdie \Cgap)/(\Cdie + \Cgap)$ is the total capacitance of the DBD arrangement \cite{liu_electrical_2003}.
The time $t=0$ is set at the instant the applied voltage $\Uapp(t)$ crosses zero and the positive half-period (HP) starts (Fig.~\ref{fig:eec-gap-voltage}).

The dynamics of the voltage across the gap (Fig.~\ref{fig:eec-gap-voltage}) provide enough information to recast the EC and obtain expressions for the discharge ignition times.
In particular, we consider the case for a DBD arrangement driven by a sinusoidal high-voltage waveform $\Uapp(t)=U_0\sin(\omega t)$ with angular frequency $\omega = 2\pi f$.
The argument relies on four assumptions: (i) $\Idisch$ is negligible when no filament is present across the gap, (ii) $\Ugap$ is approximately equal to the breakdown voltage $\Ubr$ shortly before a discharge, (iii) $\Ugap$ collapses to zero immediately after a discharge, and (iv) the DBD is in a quasi-periodic state.

Starting at the time $t=t_\mathrm{g}$ when $\Ugap$ crosses zero up until $t=t_1$ shortly before the discharge event in the positive HP, no discharges have taken place, the discharge current and its integral over time are both negligible.
At the instant $t_1$, $U_g$ is approximately equal to the breakdown voltage $\Ubr$, and Eq.~\eqref{eq:eec-gap-voltage} simplifies to 
\begin{equation}
    \Ubr \approx \Ugap(t_1) = \frac{\CT}{\Cgap}\Uapp(t_1) + \Ugap(t=0). \label{eq:t1-eec-gap-voltage}
\end{equation}

Immediately after the discharge event in the negative HP ($t=t_2$), the voltage across the gap collapses.
Since both discharges have now taken place, the integral of $\Idisch(t)$ is the sum of the charge transferred during both discharge events
\begin{equation}
    \int^{t_2}_{t_\mathrm{g}}\Idisch(\tau)\,\dd \tau = Q_1 + Q_2,
\end{equation}
assuming that the DBD is in a quasi-periodic state, i.e., $\Ugap(0) \approx \Ugap(1/f)$, implies that $Q_1 \approx -Q_2$ and Eq.~\eqref{eq:eec-gap-voltage} simplifies to
\begin{equation}
    0 \approx \Ugap(t_2) = \frac{\CT}{\Cgap}\Uapp(t_2) + \Ugap(t=0). \label{eq:t2-eec-gap-voltage}
\end{equation}

Since the waveform of the applied voltage is known, it can be substituted in \eqref{eq:t1-eec-gap-voltage} and \eqref{eq:t2-eec-gap-voltage}, and inverted to obtain equations that approximate the ignition times $t_1$ and $t_2$
\begin{align}
    t_1 &= \frac{1}{\omega} \arcsin\left[ \frac{\Cgap}{\CT U_0} \left( \Ubr -  \Ugap(t=0)\right)\right],\label{eq:t1-eec}\\[5pt]
    t_2 &= \frac{\pi}{\omega} - \frac{1}{\omega}\arcsin\left[ -\frac{\Cgap\Ugap(t=0)}{\CT U_0} \right].\label{eq:t2-eec}
\end{align}
These equations relate the effective capacitances, the applied voltage waveform, and the initial condition for the gap voltage $\Ugap(t=0)$ to the ignition times of filamentary discharges.
Depending on the information sought after, the Eqs.~\eqref{eq:t1-eec} and \eqref{eq:t2-eec} can be combined to predict the ignition time $t_2$ in the negative HP from the ignition time $t_1$ in the preceding positive HP,
\begin{equation}
    t_2(t_1) = \frac{\pi}{\omega} - \frac{1}{\omega}\arcsin\left[\sin\left(\omega t_1\right) - \frac{\Cgap \Ubr}{\CT U_0} \right].
    \label{model}
\end{equation}
Otherwise, if both ignition times $t_1$ and $t_2$ are available from a inception event during the positive HP directly followed by a inception event during the negative HP; then the Eqs.~\eqref{eq:t1-eec} and \eqref{eq:t2-eec} can be used to estimate the value of the voltage across the gap during breakdown $\Ubr$, i.e.
\begin{equation}
    \Ubr(t_1, t_2) = \frac{\CT U_0}{\Cgap}\left[ \sin(\omega t_1) - \sin(\omega t_2) \right]. \label{eq:eec-breakdown-voltage}
\end{equation}

The Eqs.~(\ref{eq:t1-eec})--(\ref{eq:eec-breakdown-voltage}) were derived under the assumptions (i)--(iv).
They predict the discharge re-ignition time $t_2$ from the ignition time $t_1$ of a discharge occurring at the same spatial position in the DBD arrangement during the preceding HP.
Despite these assumptions, it will be shown later that the model allows more general conclusions.
\begin{figure*}
    \centering
    \includegraphics[width=0.95\textwidth]{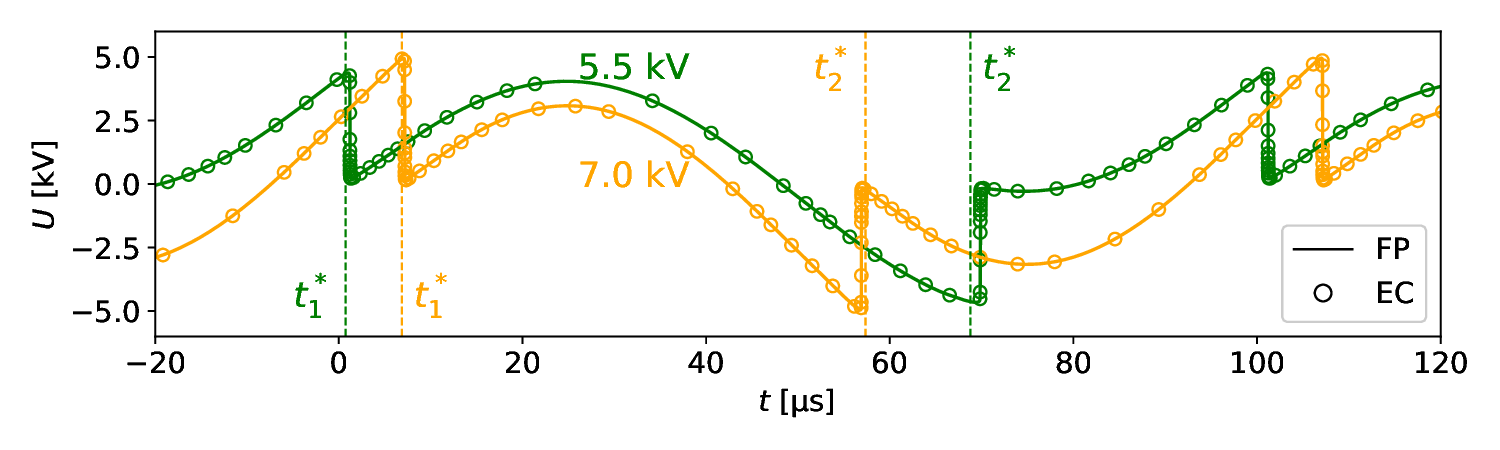}
    \caption{Validation of the EC using the results of an FP model. Two cases for the applied voltage amplitude $U_0$ were considered: 5.5\,kV (in green) and 7.0\,kV (in orange). The continuous lines show the results of the FP model, and the markers the validation of the EC. The dashed vertical lines indicate the predictions $t^*_1$ and $t^*_2$ of the ignition times from the EC model.}
    \label{fig:validation-ec-with-fp}
\end{figure*}
The results from experimental and simulation campaigns are used to validate the derived predictions for the ignition times~\eqref{eq:t1-eec} and \eqref{eq:t2-eec}. 
For the simulation campaign, the problem was simplified to a spatially 1D parallel-plate configuration with a 1\,mm gas gap and two 1 mm-thick alumina dielectric barriers (permittivity $\varepsilon_\mathrm{d}=9\varepsilon_0$).
The MCPlas toolbox \cite{stankov_mcplas_2026} was used to setup an FP model with a reaction kinetic scheme for dry synthetic air with 10~species and 58~reactions in the central layer.
The powered electrode was driven by a sinusoidal voltage with a frequency of 10\,kHz, and an amplitude $U_0$ of 5.5\,kV and 7.0\,kV.
The FP model is self-consistent, i.e., it allows the calculation of the discharge current $\Idisch(t)$ and the voltage across the gap $\Ugap(t)$.
To perform the validation, the effective capacitances are given by $\Cgap=\varepsilon_\mathrm{air} A/d$ for the gap, and $\Cgap=\varepsilon_\mathrm{d} A/d$ for each dielectric. The area $A$ is kept constant and cancels out of expressions Eq.~\eqref{eq:eec-gap-voltage}, ~\eqref{eq:t1-eec} and~\eqref{eq:t2-eec}.
The data and parameter values are then substituted in Eq.~\eqref{eq:eec-gap-voltage} to corroborate if it is possible to incorporate the dynamics of a double sided DBD arrangement in the single term $\Cdie$, obtaining a satisfactory agreement shown in Fig.~\ref{fig:validation-ec-with-fp}.
To validate the ignition times predictions~\eqref{eq:t1-eec} and~\eqref{eq:t2-eec}, the breakdown voltage $\Ubr$ was approximated as the maximum value of $\Ugap(t)$ before the first discharge. The results of the FP model are in close agreement with the expressions for both ignition times.

\begin{figure*}
    %\centering
    \includegraphics[width=0.85\textwidth]{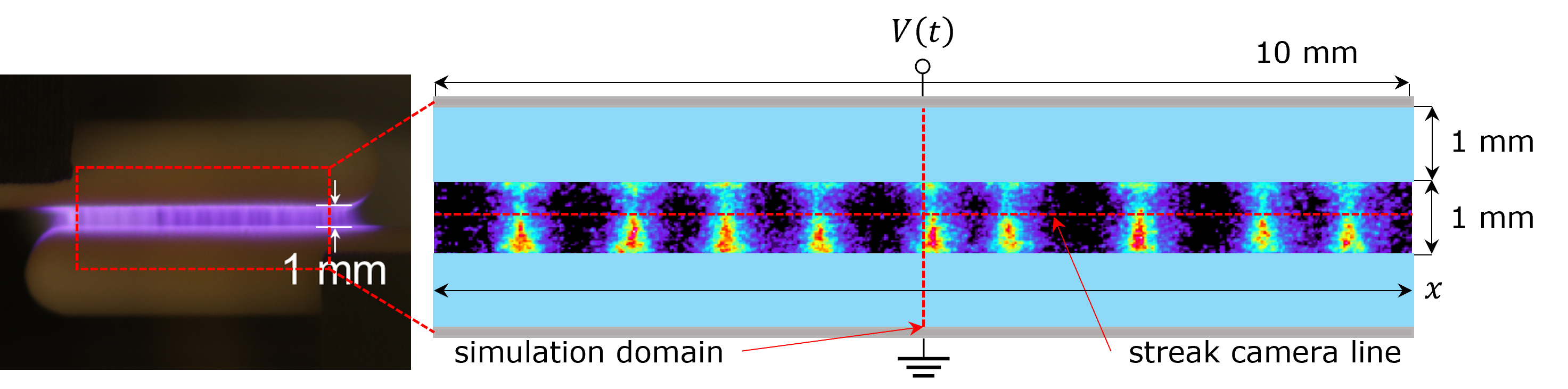}
    \caption{Detail of the multi-filament DBD arrangement: on the left, a photograph of the arrangement during operation; on the right, a diagram showing the dimensions and materials (dielectrics in blue and metal electrodes in grey), with an iCCD image exemplifying the spatial distribution of the filamentary discharges (single shot showing the spatial distribution during one positive HP). The red dashed vertical line in the diagram indicates the region simulated by the FP model, the domain is simplified to a 1D parallel plate geometry. The streak camera captures a 10\,mm line along the gap in axial centre (indicated by the dashed horizontal red line) parallel to the $x$-axis.}
    \label{fig:experimental-arrangement}
\end{figure*}
\begin{figure*}
    \centering
    \begin{subfigure}{0.48\textwidth}
        \centering
        \includegraphics[width=\textwidth]{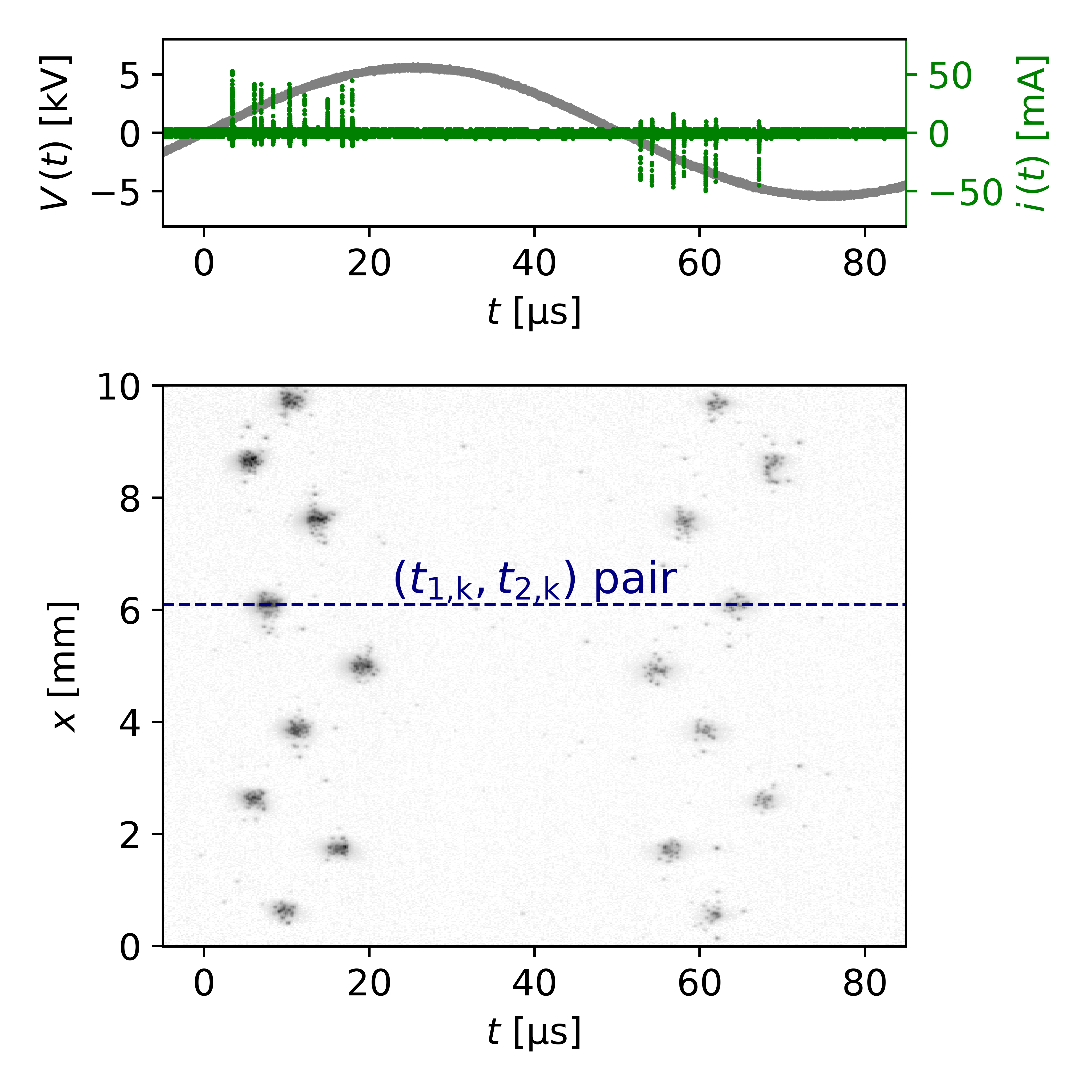}
        \caption{$U_0=5.5$\,kV}
        \label{fig:exp-data-11kVpp}
    \end{subfigure}%
    ~
    \begin{subfigure}{0.48\textwidth}
        \centering
        \includegraphics[width=\textwidth]{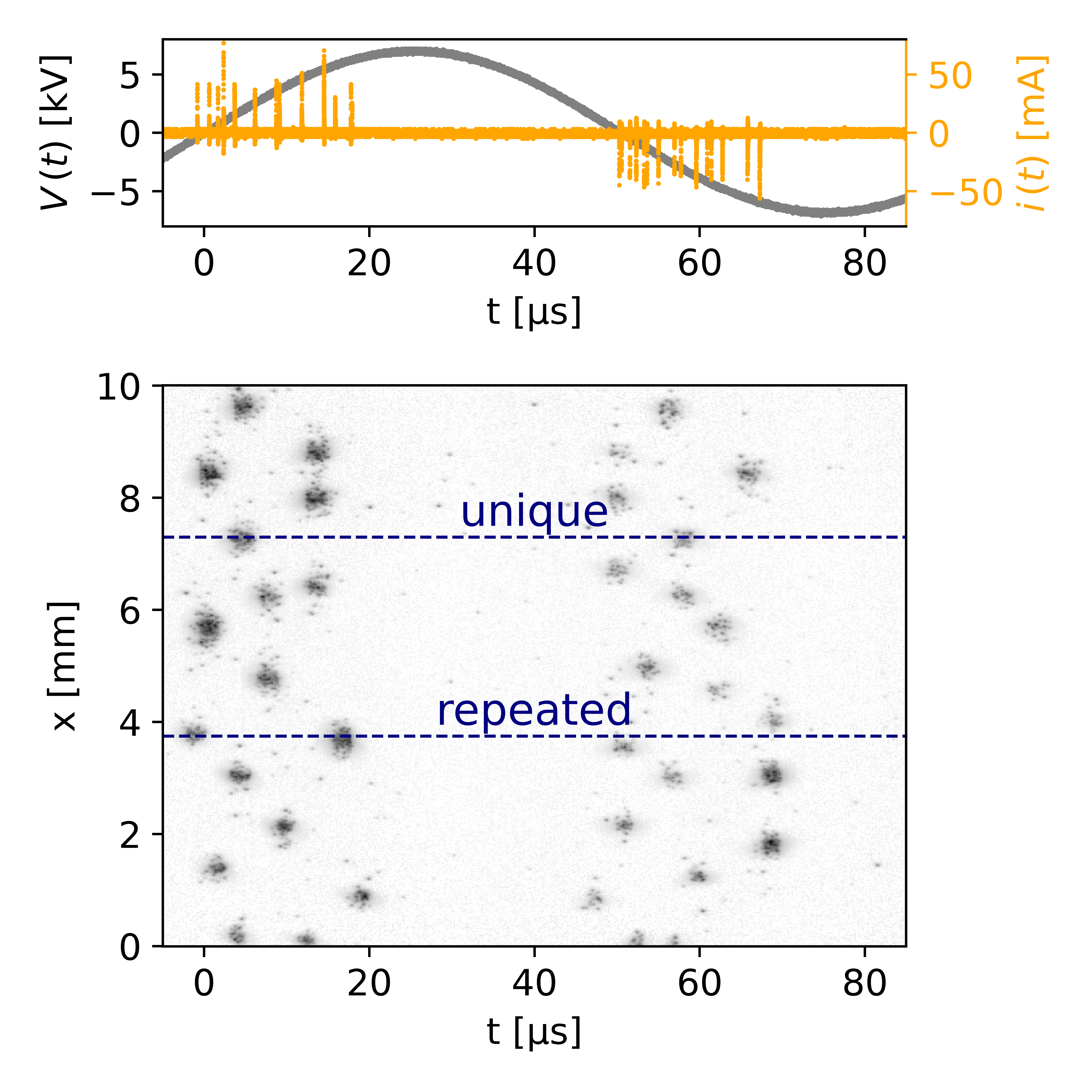}
        \caption{$U_0=7.0$\,kV}
        \label{fig:exp-data-14kVpp}
    \end{subfigure}%
    \caption{Examples of the applied voltage, total current and streak data measured during a single period of $\Uapp(t)$ with frequency $f=10$\,kHz and amplitude $U_0$ of (a) 5.5\,kV and (b) 7.0\,kV. The black spots in the streak image correspond to filamentary discharges inside the DBD. In the 7.0\,kV case, repeated discharges can take place at the same position along the arrangement during the same positive or negative HP.}
    \label{fig:experimental-data-instances}
\end{figure*}

Following up on the work done in \cite{guikema_spontaneous_2000} and \cite{klein_time-resolved_2001}, a double-sided symmetric DBD arrangement with an effectively 1D lateral geometry was selected as the experimental setup (Fig.~\ref{fig:experimental-arrangement}).
The details of the DBD arrangement, as well as the diagnostic setup are the same as in \cite{hoft_upscaling_2022}.
This geometry constrains the filaments to a single plane parallel to the applied electric field and allows to completely specify the location of each filament inside the arrangement using a single space coordinate.
The arrangement consists of two cylindrical electrodes covered by a layer of alumina 1\,mm thick.
The gap between the dielectrics has also a thickness of 1\,mm.
A gas mixture of dry synthetic air (20\,vol\% O\textsubscript{2} in N\textsubscript{2}) at atmospheric pressure was used and the DBD was driven by a sinusoidal waveform with a frequency of 10\,kHz.
The total current and the applied voltage were measured, and a streak camera pointing towards the gap measured the spatial and temporal distribution of the filamentary discharges.
Two sets of measurements were recorded with different applied voltage amplitudes $U_0$, the first for 5.5\,kV, and the second one for 7.0\,kV (Fig.~\ref{fig:experimental-data-instances}).

\begin{figure*}
    \centering
    \begin{subfigure}{0.41\textwidth}
        \centering
        \vspace{1mm}
        \includegraphics[width=\textwidth]{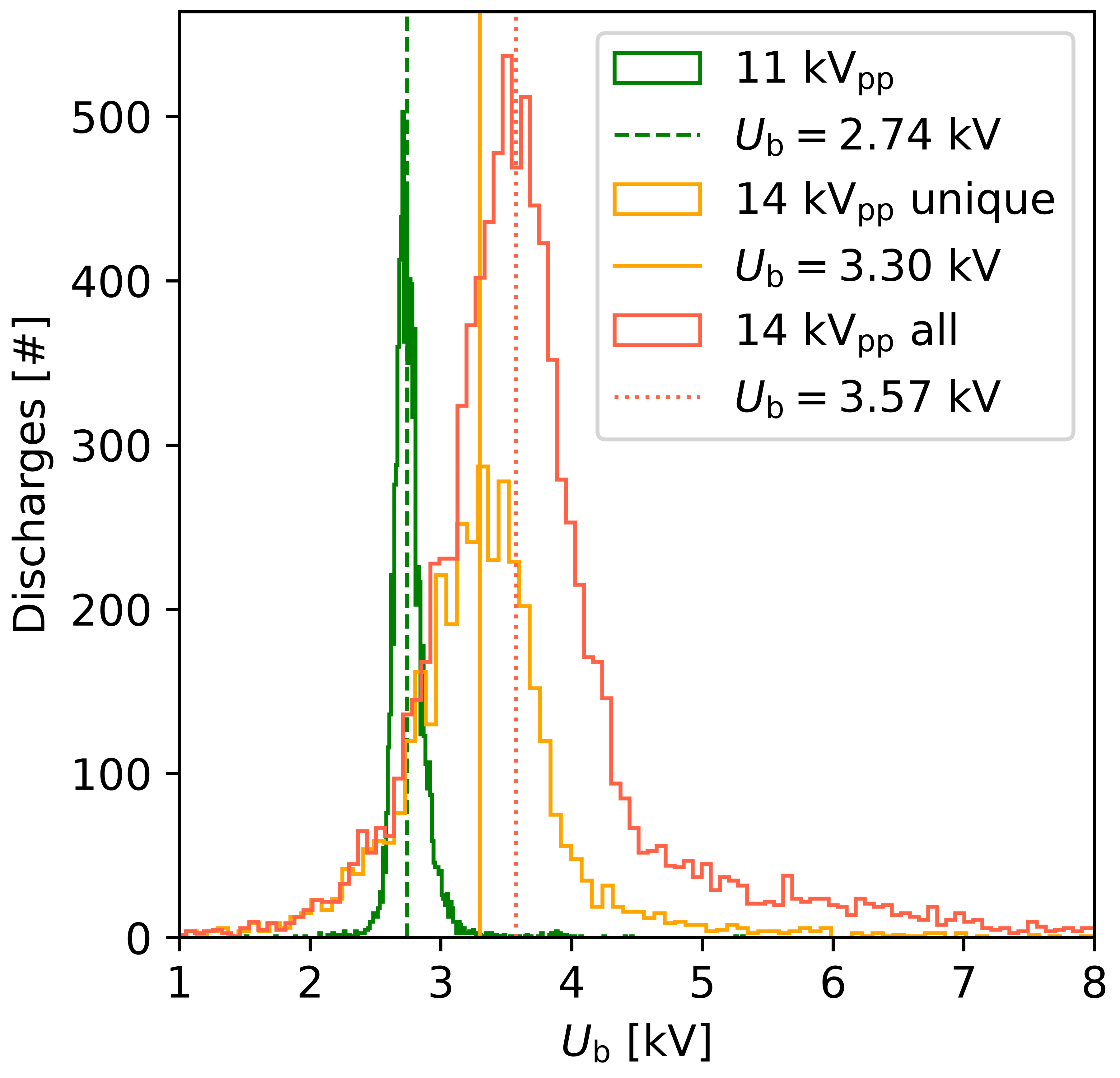}
        %\vspace{0.5mm}
        \caption{}%$U_b$ from experimental $(t_{1,k}, t_{2,k})$ pairs
        \label{fig:breakdown-voltage-histogram}
    \end{subfigure}%
    \hspace{\fill}
    \begin{subfigure}{0.49\textwidth}
        \centering
        \includegraphics[width=\textwidth]{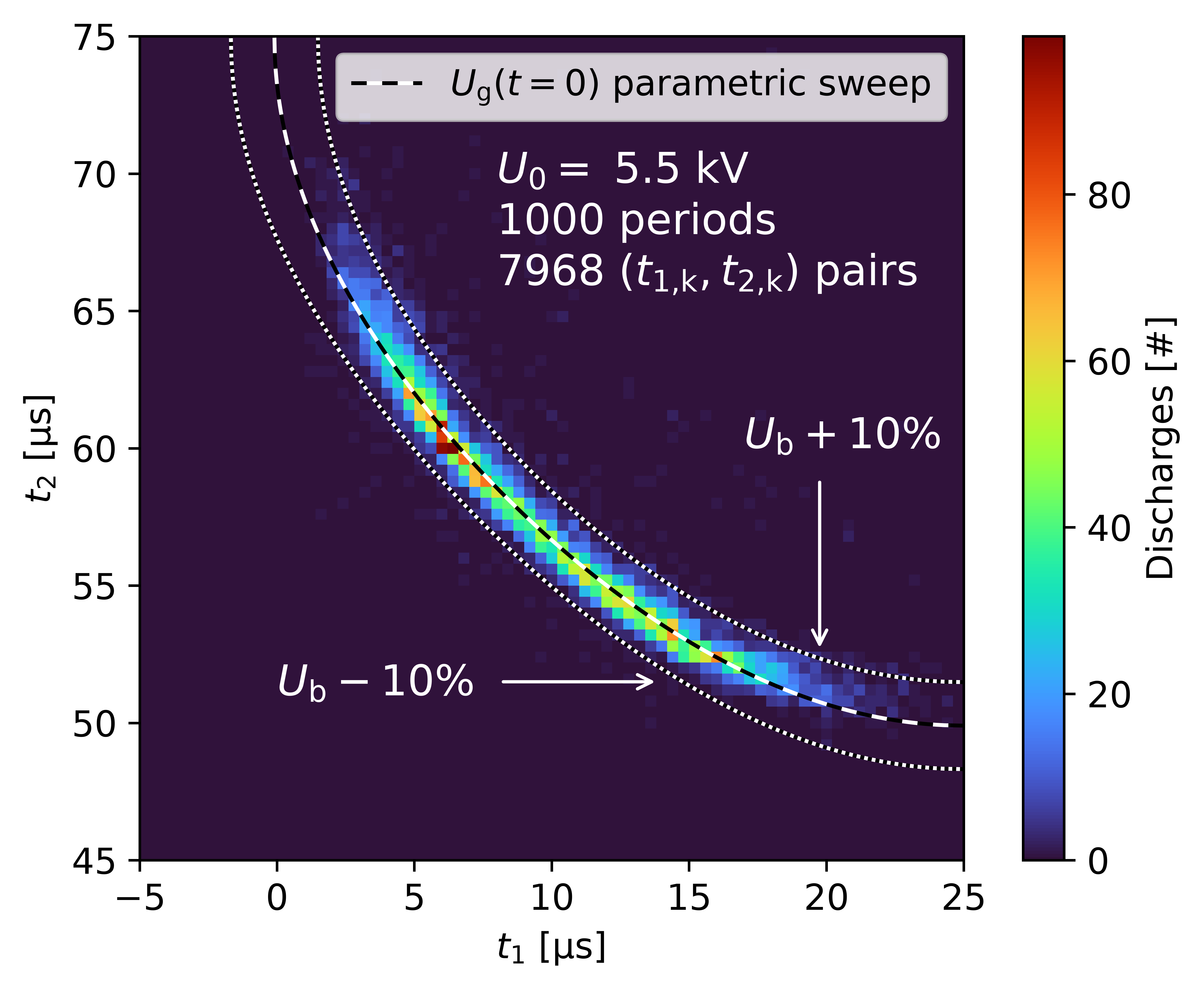}
        \caption{} %$U_0=5.5$\,kV$U_0=5.5$\,kV
        \label{fig:time-correlation-11kVpp}
    \end{subfigure}

    \begin{subfigure}{0.49\textwidth}
        \centering
        \includegraphics[width=\textwidth]{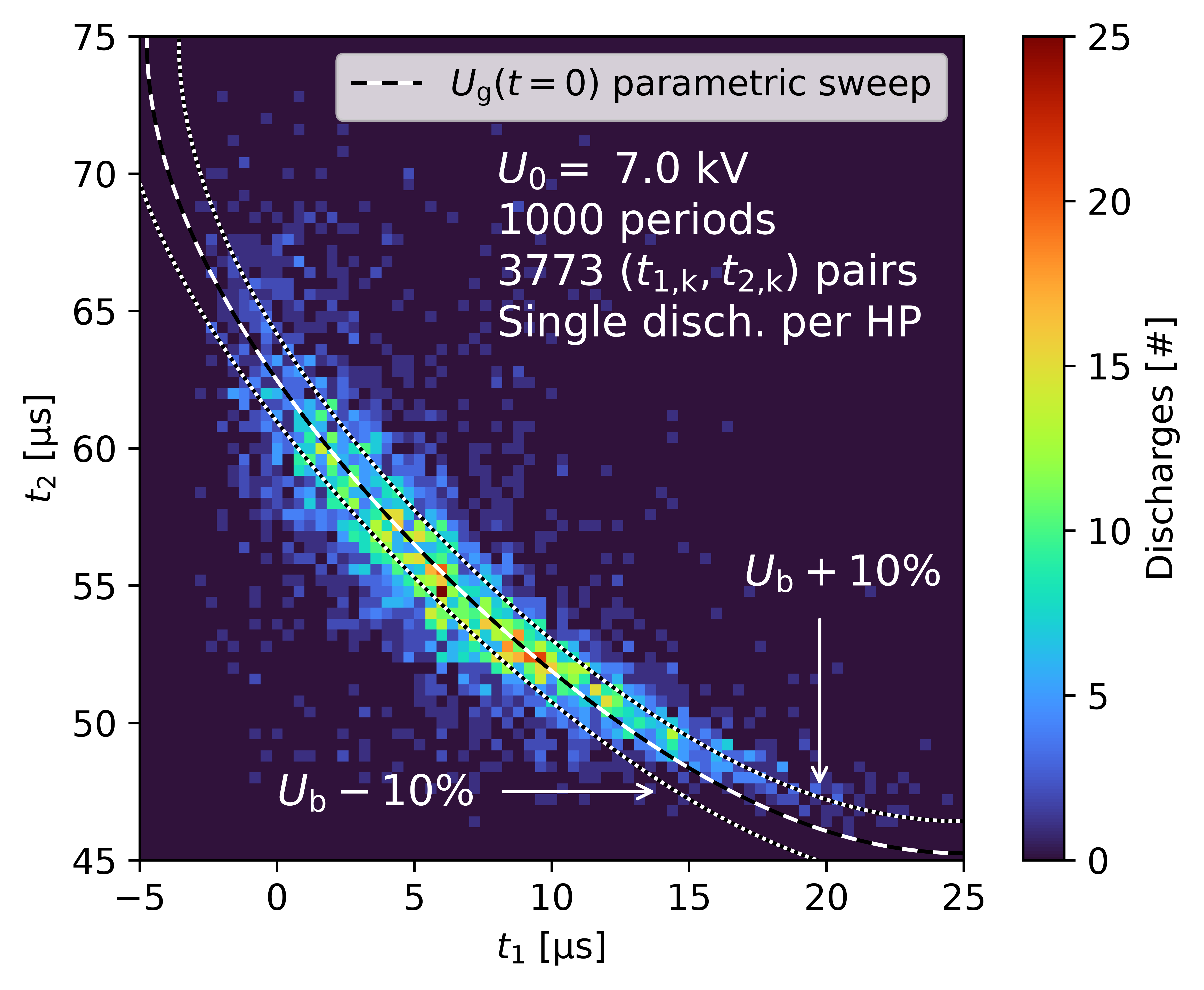}
        \caption{} %$U_0=7.0$\,kV, unique discharges per HP
        \label{fig:time-correlation-14kVpp-unique}
    \end{subfigure}%
    \hspace{\fill}
    \begin{subfigure}{0.49\textwidth}
        \centering
        \includegraphics[width=\textwidth]{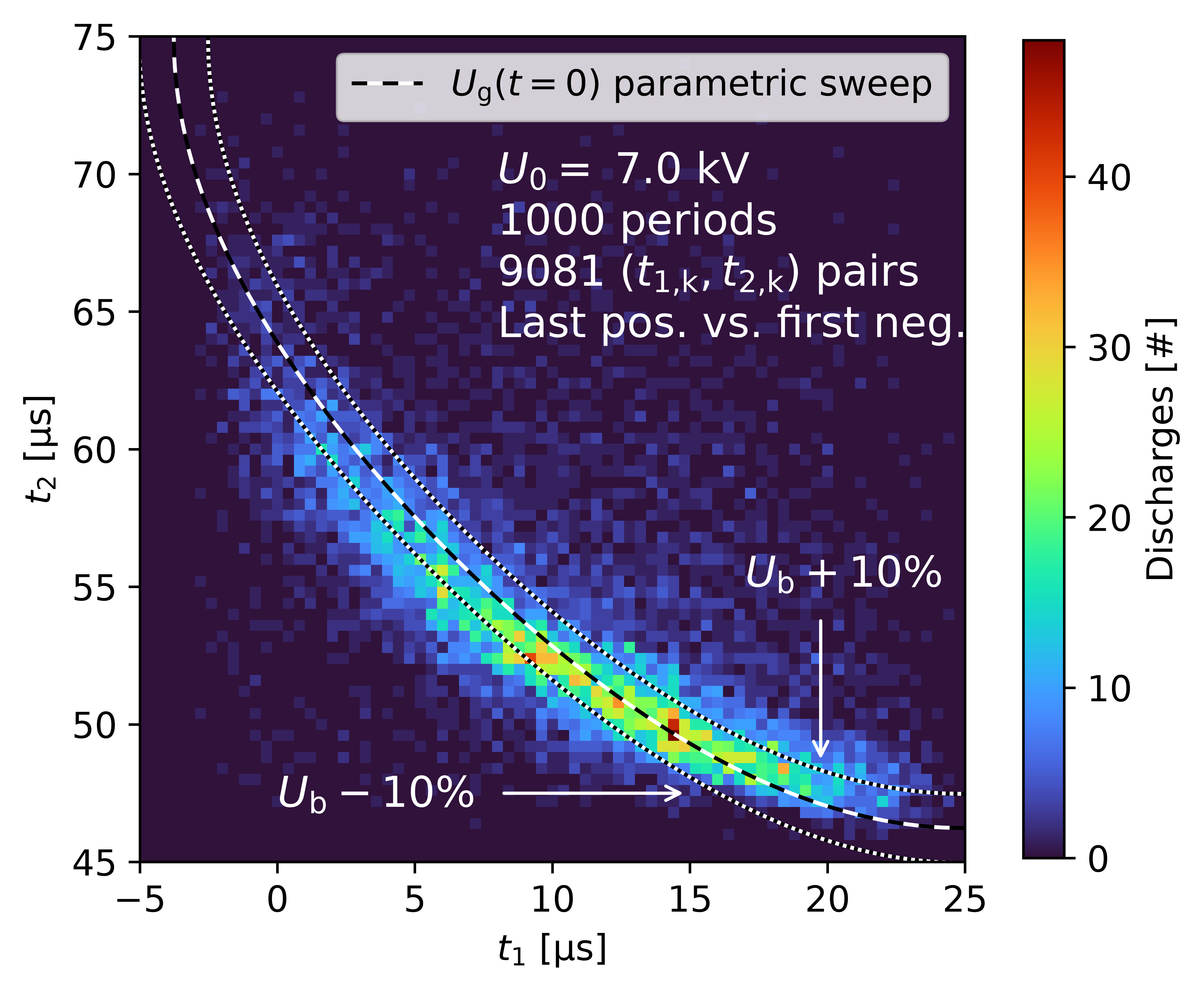}
        \caption{} %$U_0=7.0$\,kV, unique and multiple discharges per HP.
        \label{fig:time-correlation-14kVpp-all}
    \end{subfigure}%
    \caption{(a) Estimations of the breakdown voltage from the ignition times measured by the streak camera.
    \mbox{(b) Distribution} of the ignition time pairs $(t_{1,k}, t_{2,k})$ from the 5.5\,kV case. (c) Results from the 7.0\,kV case with the ignition time pairs from unique discharges per HP separated from neighbouring discharges by at least 0.57\,mm.
    \mbox{(d) Distribution} of the ignition time pairs from the 7.0\,kV case with the unique discharges per HP, and, when multiple discharges are present, the last discharge event in the positive HP and the first discharge event in the negative HP are selected.}
    \label{fig:validation-with-experiment}
\end{figure*}

For the applied voltage amplitude $U_0=5.5$\,kV, the streak images show discharges appearing at regular intervals in space and time creating a highly ordered pattern stable over several periods (Fig.~\ref{fig:exp-data-11kVpp}).
An average of 16 filamentary discharges occur during a complete period of the applied voltage.
The median inter-filament distance is 1.1\,mm, and 98\% of the filaments are separated by more than 0.75\,mm.
A discharge occurring during the positive HP of the applied voltage is always followed by a discharge during the negative HP at the same position along the 1D  arrangement.
At this applied voltage amplitude, no subsequent discharges appear at the same position during the same HP, the deposited charge at the dielectrics' surface precludes repeated ignitions for the same polarity of $\Uapp(t)$.
It is then possible to pair a unique discharge appearing during the positive HP and at position $x_k$ and ignition time $t_{1,k}$ with a unique counterpart igniting at the same position at a later time $t_{2,k}$ during the negative HP.
The ignition time pairs $(t_{1,k}, t_{2,k})$ present a correlation where earlier discharges during the positive HP correspond to later discharges during the negative HP and vice versa, as shown in Fig.~\ref{fig:time-correlation-11kVpp}.


The Eqs.~\eqref{eq:t1-eec} and~\eqref{eq:t2-eec} can be validated with experimental data by means of a parametric sweep over the initial condition $\Ugap(t=0)$, linking the local gap voltage with the ignition times.
First, the effective capacitances are calculated from the average Q-V curve \cite{pipa_simplest_2012}, obtaining $\Cdie=0.8$\,pF and $\CT=0.4$\,pF.
These values correspond to the entire multi-filament arrangement.
It has been shown previously that the electrical quantities from a single-filament arrangement are proportional to the number of discharges present in a multi-filament setup \cite{hoft_upscaling_2022}.
Thus, a scaling factor must be used when analysing single discharges occurring in a multi-filament setup.
However, we can avoid determining such factor because the Eqs.~\eqref{eq:t1-eec}, \eqref{eq:t2-eec}, and \eqref{eq:eec-breakdown-voltage} include only the ratios of the effective capacitances where the scaling effect cancels out.
This fact broadens the applicability of the model from a zero-dimensional description of a single filament to a general description of multi-filament DBDs, provided that exactly one discharge ignites at a given position during the same HP, such cases are depicted in Fig.~\ref{fig:exp-data-11kVpp}.
Once the values of the effective capacitances have been determined, the Eq.~\eqref{eq:eec-breakdown-voltage} and the built $(t_{1,k}, t_{2,k})$ pairs are used to calculate the median breakdown voltage $\Ubr=2.74$\,kV (Fig.~\ref{fig:breakdown-voltage-histogram}).
Finally, the parametric sweep over $\Ugap(t=0)$ is performed and its results describe the central tendency of the correlation between the re-ignition times $t_{1,k}$ and $t_{1,k}$, this is shown in Fig.~\ref{fig:time-correlation-11kVpp}.
Furthermore, the dispersion observed in the experimental data can be explained by small variations of the involved quantities.
For instance, different pre-ionisation states can result in different breakdown voltages \cite{Nemschokmichal-2018-ID5108}. The bounds obtained from varying the value of $\Ubr$ by $\pm10$\,\% include 96\,\% of the $(t_{1,k}, t_{2,k})$ pairs.
If the measurement uncertainties of the effective capacitances are taken into account, similar bounds are obtained.

The spatio-temporal pattern that emerges with an applied voltage amplitude $U_0=7.0$\,kV shows several differences to the 5.5\,kV case (Fig.~\ref{fig:exp-data-14kVpp}).
A larger number of filaments appear during a complete period, with an average of 36.
The median inter-filament distance is of only 0.57\,mm, and there is not a clear minimum separation for the majority of discharges.
In contrast to the 5.5\,kV case, there are repeated discharges occurring at the same position and HP. The surface charge distribution becomes more complex and so do the dynamics of the system, slowly entering into a more unordered regime.

Firstly, to evaluate the validity of the ignition time predictions, only the cases with a single discharge at the same position during each HP and an inter-filament distance larger than 0.57\,mm on both sides are considered.
The latter constraint ensures that the filaments are as isolated as possible from the effects of neighbouring discharges.
The validation of the ignition times predictions proceeds in a similar fashion to the 5.5\,kV case.
The effective capacitances obtained from the Q-V curve are $\Cdie=1.4$\,pF and $\CT=0.4$\,pF.
The median breakdown voltage obtained from Eq.~\eqref{eq:eec-breakdown-voltage} is $\Ubr=3.3$\,kV.
However, the values show a greater dispersion when compared to the distribution from the lower voltage, see Fig.~\ref{fig:breakdown-voltage-histogram}.
This larger dispersion is also present in the 2D histogram of the $(t_{1,k}, t_{2,k})$ pairs, see  Fig.~\ref{fig:time-correlation-14kVpp-unique}.
Nevertheless, a similar correlation is present among the ignition times, and the Eqs.~\eqref{eq:t1-eec} and~\eqref{eq:t2-eec} can still describe its central tendency, and the bounds obtained after varying the breakdown voltage still include 53\,\% of the ignition times pairs.

Secondly, the validity of the predicted ignition times can also be evaluated for cases in which multiple discharges occur during a single HP.
In this case, however, $t_1$ must be taken as the ignition time of the last discharge in the half-period, whereas $t_2$ remains the ignition time of the first discharge in the subsequent half-period.
Only under these conditions is the assumption (iv) satisfied, implying a zero net charge balance over the two consecutive HPs.
Again, a similar correlation is present among the ignition times, as for the previous cases, and the Eqs.~\eqref{eq:t1-eec} and~\eqref{eq:t2-eec} can still describe its central tendency (shown in Fig.\,\ref{fig:time-correlation-14kVpp-all}), although the distribution shows a larger dispersion.
This can also be observed in the long tail towards higher breakdown voltages in Fig.~\ref{fig:breakdown-voltage-histogram}.
The bounds obtained after varying the breakdown voltage still include 50\,\% of the ignition times pairs.
Apparently, the dispersion of the data around the central correlation curve predicted by the model contains implicit information about the regularity, or degree of order, of the system. The greater the dispersion in the re-ignition-time correlation, the weaker is the spatio-temporal order of the investigated system.

\begin{figure}[ht]
    \centering
    \includegraphics[width=0.49\textwidth]{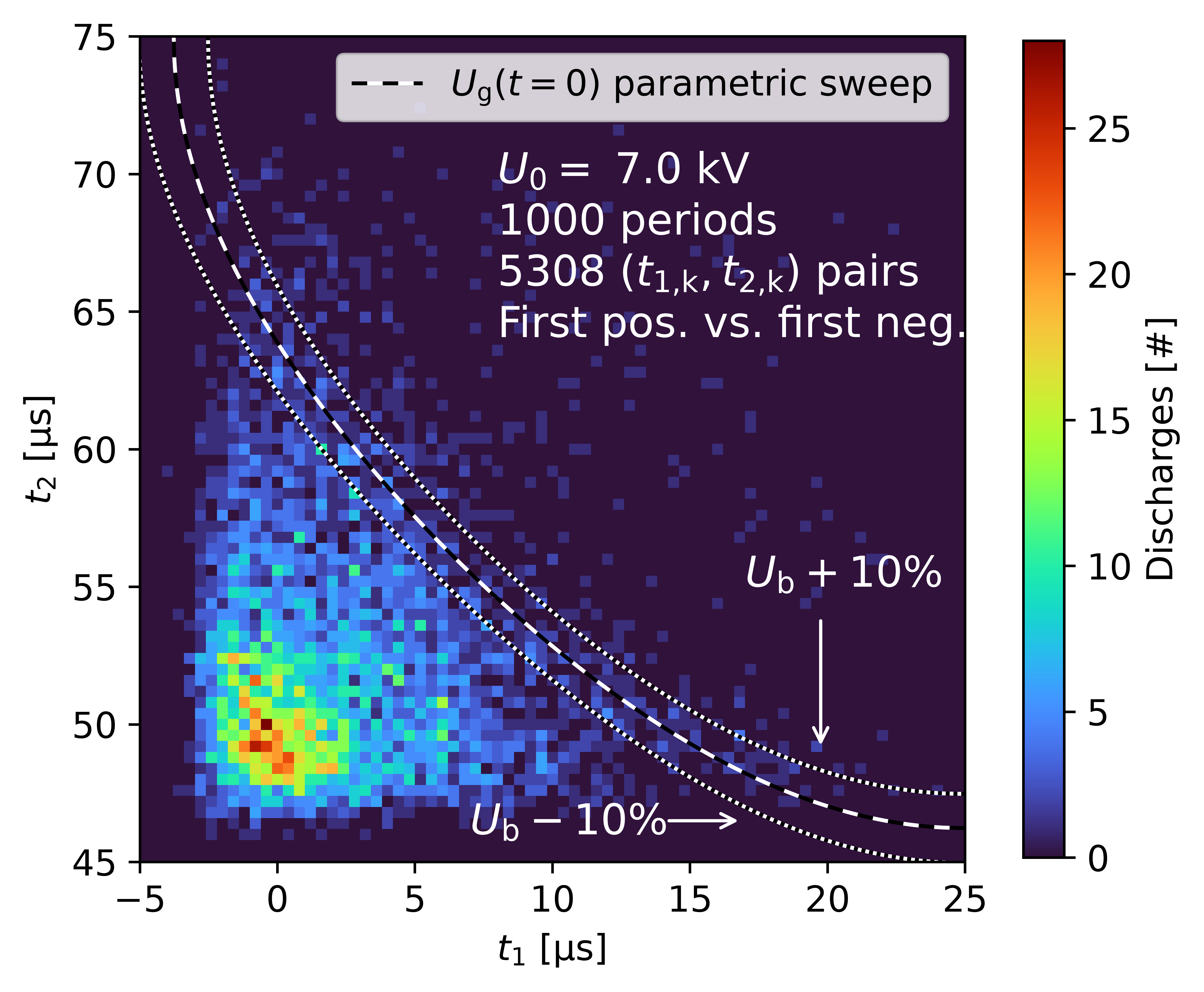}
    \caption{Distribution of the ignition time pairs $(t_{1,k}, t_{2,k})$ from the 7.0\,kV case. Here, the pairs were build taking the first discharge events in both HPs only for the cases where multiple events occurs in the same HP. It can be seen that they cluster around the start of their respective HP.}
    \label{fig:all-14kVpp-discharges-first-positive-first-negative}
\end{figure}

The ignition patterns for the 7.0\,kV case show a greater variety of dynamics, where multiple discharges can occur in either the positive or negative HP at the same position along the arrangement (Fig.~\ref{fig:exp-data-14kVpp}).
The discharge ignition pairs $(t_{1,k}, t_{2,k})$ shown in Fig.~\ref{fig:time-correlation-14kVpp-all} were built by selecting the last discharge event in the positive HP and the first discharge event in the negative HP to remove any effect of intermediate discharges, which invalidate assumption (iv).
Any other selection strategy has to link ignition times pairs of nonconsecutive discharges.
Pairing, for instance, the first discharge event in the positive HP with the first discharge event during the negative HP yields a different correlation (Fig.~\ref{fig:all-14kVpp-discharges-first-positive-first-negative}).
In this case, the ignition times cluster around the start of each HP, away from the central tendency observed in Fig.~\ref{fig:time-correlation-14kVpp-all} for discharge pairs built from the last ignition in the positive HP and first ignition in the negative HP.
The behaviour between the first discharge event of the positive HP and the first discharge event of the negative HP is not considered by the assumptions (i)--(iv); inside the time interval $[t_{1,k}, t_{2,k}]$ the local electric potential has both reached the breakdown value and collapsed due to the intermediate discharge, and the charge balance is not met due to the additional discharge current peak.

In conclusion, the reduced-order model (Eqs.~\eqref{eq:t1-eec} and~\eqref{eq:t2-eec}) derived in this work can predict the ignition times for filamentary discharges igniting in a DBD driven by a sinusoidal high voltage.
The validity of this model was verified computationally for a parallel plate arrangement by means of a fluid-Poisson model, and experimentally with the measurements conducted in a multi-filamentary DBD arrangement in dry synthetic air at atmospheric pressure with an effective 1D lateral geometry.
In both studies, the voltage amplitudes 5.5\,kV and 7.0\,kV were investigated.
The difference in the dynamics between both cases are clearly illustrated in the experimental results, where the 5.5\,kV amplitude generates a regular spatio-temporal pattern, the stability of which is perturbed at higher voltages.
Nevertheless, the central tendency and bounds provided by the Eqs.~\eqref{eq:t1-eec} and~\eqref{eq:t2-eec} can reliably describe the non-linear correlation between the ignition times of filamentary discharges.
Ongoing work focuses on the dynamics of multiple discharges per HP- (Fig.~\ref{fig:all-14kVpp-discharges-first-positive-first-negative})  and the synthesis of experimental and simulation results to formulate a spatially resolved reduced-order model.

\section*{Funding}
This work was funded by the Deutsche Forschungsgemeinschaft (DFG, German Research Foundation)---project number~535827833.
The contribution of TH was supported by project~LM2023039 funded by the Ministry of Education, Youth and Sports of the Czech Republic.

\bibliography{references}
\end{document}

%% file: eec_diagram.tikz
\begin{tikzpicture}[transform shape]
	% Paths, nodes and wires:
	\draw (7, 5) to[capacitor, l_={$\Cdie$}] (8, 5);
	\draw (8.5, 5) to[capacitor, l_={$\Cgap$}] (9.5, 5);
	\draw (8.5, 6.25) to[american current source, l={$\Idisch(t)$}] (9.5, 6.25);
	\draw (6.375, 5) -- (7, 5);
	\draw (8, 5) -- (8.5, 5);
	\draw (8.5, 6.25) -- (8.25, 6.25) -| (8.25, 5);
	\draw (9.5, 6.25) -- (9.75, 6.25) -| (9.75, 5);
	\node[ground] at (10, 5){};
	\draw (9.5, 5) -- (10, 5);
	%\node[circ](N1) at (10, 5){} node[anchor=west] at (N1.east){$\Itotal(t)$};
	\node[ocirc](N2) at (6.375, 5){} node[anchor=south] at (N2.north){$\Uapp(t)$};
	\node[circ] at (8.25, 5){};
	\node[circ] at (9.75, 5){};
\end{tikzpicture}

%% file: main_text.bbl
\begin{thebibliography}{28}%
\makeatletter
\providecommand \@ifxundefined [1]{%
 \@ifx{#1\undefined}
}%
\providecommand \@ifnum [1]{%
 \ifnum #1\expandafter \@firstoftwo
 \else \expandafter \@secondoftwo
 \fi
}%
\providecommand \@ifx [1]{%
 \ifx #1\expandafter \@firstoftwo
 \else \expandafter \@secondoftwo
 \fi
}%
\providecommand \natexlab [1]{#1}%
\providecommand \enquote  [1]{``#1''}%
\providecommand \bibnamefont  [1]{#1}%
\providecommand \bibfnamefont [1]{#1}%
\providecommand \citenamefont [1]{#1}%
\providecommand \href@noop [0]{\@secondoftwo}%
\providecommand \href [0]{\begingroup \@sanitize@url \@href}%
\providecommand \@href[1]{\@@startlink{#1}\@@href}%
\providecommand \@@href[1]{\endgroup#1\@@endlink}%
\providecommand \@sanitize@url [0]{\catcode `\\12\catcode `\$12\catcode
  `\&12\catcode `\#12\catcode `\^12\catcode `\_12\catcode `\%12\relax}%
\providecommand \@@startlink[1]{}%
\providecommand \@@endlink[0]{}%
\providecommand \url  [0]{\begingroup\@sanitize@url \@url }%
\providecommand \@url [1]{\endgroup\@href {#1}{\urlprefix }}%
\providecommand \urlprefix  [0]{URL }%
\providecommand \Eprint [0]{\href }%
\providecommand \doibase [0]{https://doi.org/}%
\providecommand \selectlanguage [0]{\@gobble}%
\providecommand \bibinfo  [0]{\@secondoftwo}%
\providecommand \bibfield  [0]{\@secondoftwo}%
\providecommand \translation [1]{[#1]}%
\providecommand \BibitemOpen [0]{}%
\providecommand \bibitemStop [0]{}%
\providecommand \bibitemNoStop [0]{.\EOS\space}%
\providecommand \EOS [0]{\spacefactor3000\relax}%
\providecommand \BibitemShut  [1]{\csname bibitem#1\endcsname}%
\let\auto@bib@innerbib\@empty
%</preamble>
\bibitem [{\citenamefont {Purwins}\ \emph {et~al.}(2010)\citenamefont
  {Purwins}, \citenamefont {Bödeker},\ and\ \citenamefont
  {Amiranashvili}}]{purwins_dissipative_2010}%
  \BibitemOpen
  \bibfield  {author} {\bibinfo {author} {\bibfnamefont {H.-G.}\ \bibnamefont
  {Purwins}}, \bibinfo {author} {\bibfnamefont {H.}~\bibnamefont {Bödeker}},\
  and\ \bibinfo {author} {\bibfnamefont {S.}~\bibnamefont {Amiranashvili}},\
  }\bibfield  {title} {\bibinfo {title} {Dissipative solitons},\ }\href
  {https://doi.org/10.1080/00018732.2010.498228} {\bibfield  {journal}
  {\bibinfo  {journal} {Advances in Physics}\ }\textbf {\bibinfo {volume}
  {59}},\ \bibinfo {pages} {485} (\bibinfo {year} {2010})}\BibitemShut
  {NoStop}%
\bibitem [{\citenamefont {Van~Brunt}(1991)}]{brunt1991}%
  \BibitemOpen
  \bibfield  {author} {\bibinfo {author} {\bibfnamefont {R.}~\bibnamefont
  {Van~Brunt}},\ }\bibfield  {title} {\bibinfo {title} {Stochastic properties
  of partial-discharge phenomena},\ }\href {https://doi.org/10.1109/14.99099}
  {\bibfield  {journal} {\bibinfo  {journal} {IEEE Transactions on Electrical
  Insulation}\ }\textbf {\bibinfo {volume} {26}},\ \bibinfo {pages} {902}
  (\bibinfo {year} {1991})}\BibitemShut {NoStop}%
\bibitem [{\citenamefont {Heitz}(1999)}]{heitz1999}%
  \BibitemOpen
  \bibfield  {author} {\bibinfo {author} {\bibfnamefont {C.}~\bibnamefont
  {Heitz}},\ }\bibfield  {title} {\bibinfo {title} {A generalized model for
  partial discharge processes based on a stochastic process approach},\ }\href
  {https://doi.org/10.1088/0022-3727/32/9/312} {\bibfield  {journal} {\bibinfo
  {journal} {Journal of Physics D: Applied Physics}\ }\textbf {\bibinfo
  {volume} {32}},\ \bibinfo {pages} {1012} (\bibinfo {year}
  {1999})}\BibitemShut {NoStop}%
\bibitem [{\citenamefont {Stollenwerk}\ \emph {et~al.}(2006)\citenamefont
  {Stollenwerk}, \citenamefont {Amiranashvili}, \citenamefont {Boeuf},\ and\
  \citenamefont {Purwins}}]{stollen2006}%
  \BibitemOpen
  \bibfield  {author} {\bibinfo {author} {\bibfnamefont {L.}~\bibnamefont
  {Stollenwerk}}, \bibinfo {author} {\bibfnamefont {S.}~\bibnamefont
  {Amiranashvili}}, \bibinfo {author} {\bibfnamefont {J.-P.}\ \bibnamefont
  {Boeuf}},\ and\ \bibinfo {author} {\bibfnamefont {H.-G.}\ \bibnamefont
  {Purwins}},\ }\bibfield  {title} {\bibinfo {title} {Measurement and 3{D}
  simulation of self-organized filaments in a barrier discharge},\ }\href
  {https://doi.org/10.1103/PhysRevLett.96.255001} {\bibfield  {journal}
  {\bibinfo  {journal} {Phys. Rev. Lett.}\ }\textbf {\bibinfo {volume} {96}},\
  \bibinfo {pages} {255001} (\bibinfo {year} {2006})}\BibitemShut {NoStop}%
\bibitem [{\citenamefont {Wild}\ and\ \citenamefont
  {Stollenwerk}(2012)}]{wild2012}%
  \BibitemOpen
  \bibfield  {author} {\bibinfo {author} {\bibfnamefont {R.}~\bibnamefont
  {Wild}}\ and\ \bibinfo {author} {\bibfnamefont {L.}~\bibnamefont
  {Stollenwerk}},\ }\bibfield  {title} {\bibinfo {title} {Breakdown of order in
  a self-organised barrier discharge},\ }\href
  {https://doi.org/10.1140/epjd/e2012-30220-4} {\bibfield  {journal} {\bibinfo
  {journal} {The European Physical Journal D}\ }\textbf {\bibinfo {volume}
  {66}},\ \bibinfo {pages} {214} (\bibinfo {year} {2012})}\BibitemShut
  {NoStop}%
\bibitem [{\citenamefont {Guikema}\ \emph {et~al.}(2000)\citenamefont
  {Guikema}, \citenamefont {Miller}, \citenamefont {Niehof}, \citenamefont
  {Klein},\ and\ \citenamefont {Walhout}}]{guikema_spontaneous_2000}%
  \BibitemOpen
  \bibfield  {author} {\bibinfo {author} {\bibfnamefont {J.}~\bibnamefont
  {Guikema}}, \bibinfo {author} {\bibfnamefont {N.}~\bibnamefont {Miller}},
  \bibinfo {author} {\bibfnamefont {J.}~\bibnamefont {Niehof}}, \bibinfo
  {author} {\bibfnamefont {M.}~\bibnamefont {Klein}},\ and\ \bibinfo {author}
  {\bibfnamefont {M.}~\bibnamefont {Walhout}},\ }\bibfield  {title} {\bibinfo
  {title} {Spontaneous pattern formation in an effectively one-dimensional
  dielectric-barrier discharge system},\ }\href
  {https://doi.org/10.1103/PhysRevLett.85.3817} {\bibfield  {journal} {\bibinfo
   {journal} {Physical Review Letters}\ }\textbf {\bibinfo {volume} {85}},\
  \bibinfo {pages} {3817} (\bibinfo {year} {2000})}\BibitemShut {NoStop}%
\bibitem [{\citenamefont {Callegari}\ \emph {et~al.}(2014)\citenamefont
  {Callegari}, \citenamefont {Bernecker},\ and\ \citenamefont
  {Boeuf}}]{callegari_pattern_2014}%
  \BibitemOpen
  \bibfield  {author} {\bibinfo {author} {\bibfnamefont {T.}~\bibnamefont
  {Callegari}}, \bibinfo {author} {\bibfnamefont {B.}~\bibnamefont
  {Bernecker}},\ and\ \bibinfo {author} {\bibfnamefont {J.~P.}\ \bibnamefont
  {Boeuf}},\ }\bibfield  {title} {\bibinfo {title} {Pattern formation and
  dynamics of plasma filaments in dielectric barrier discharges},\ }\href
  {https://doi.org/10.1088/0963-0252/23/5/054003} {\bibfield  {journal}
  {\bibinfo  {journal} {Plasma Sources Science and Technology}\ }\textbf
  {\bibinfo {volume} {23}},\ \bibinfo {pages} {054003} (\bibinfo {year}
  {2014})}\BibitemShut {NoStop}%
\bibitem [{\citenamefont {Klein}\ \emph {et~al.}(2001)\citenamefont {Klein},
  \citenamefont {Miller},\ and\ \citenamefont
  {Walhout}}]{klein_time-resolved_2001}%
  \BibitemOpen
  \bibfield  {author} {\bibinfo {author} {\bibfnamefont {M.}~\bibnamefont
  {Klein}}, \bibinfo {author} {\bibfnamefont {N.}~\bibnamefont {Miller}},\ and\
  \bibinfo {author} {\bibfnamefont {M.}~\bibnamefont {Walhout}},\ }\bibfield
  {title} {\bibinfo {title} {Time-resolved imaging of spatiotemporal patterns
  in a one-dimensional dielectric-barrier discharge system},\ }\href
  {https://doi.org/10.1103/PhysRevE.64.026402} {\bibfield  {journal} {\bibinfo
  {journal} {Phys. Rev. E}\ }\textbf {\bibinfo {volume} {64}},\ \bibinfo
  {pages} {026402} (\bibinfo {year} {2001})}\BibitemShut {NoStop}%
\bibitem [{\citenamefont {Stollenwerk}\ \emph {et~al.}(2007)\citenamefont
  {Stollenwerk}, \citenamefont {Laven},\ and\ \citenamefont
  {Purwins}}]{stollen2007}%
  \BibitemOpen
  \bibfield  {author} {\bibinfo {author} {\bibfnamefont {L.}~\bibnamefont
  {Stollenwerk}}, \bibinfo {author} {\bibfnamefont {J.~G.}\ \bibnamefont
  {Laven}},\ and\ \bibinfo {author} {\bibfnamefont {H.-G.}\ \bibnamefont
  {Purwins}},\ }\bibfield  {title} {\bibinfo {title} {Spatially resolved
  surface-charge measurement in a planar dielectric-barrier discharge system},\
  }\href {https://doi.org/10.1103/PhysRevLett.98.255001} {\bibfield  {journal}
  {\bibinfo  {journal} {Phys. Rev. Lett.}\ }\textbf {\bibinfo {volume} {98}},\
  \bibinfo {pages} {255001} (\bibinfo {year} {2007})}\BibitemShut {NoStop}%
\bibitem [{\citenamefont {Höft}\ \emph {et~al.}(2014)\citenamefont {Höft},
  \citenamefont {Kettlitz}, \citenamefont {Becker}, \citenamefont {Hoder},
  \citenamefont {Loffhagen}, \citenamefont {Brandenburg},\ and\ \citenamefont
  {Weltmann}}]{Hoeft-2014-ID3403}%
  \BibitemOpen
  \bibfield  {author} {\bibinfo {author} {\bibfnamefont {H.}~\bibnamefont
  {Höft}}, \bibinfo {author} {\bibfnamefont {M.}~\bibnamefont {Kettlitz}},
  \bibinfo {author} {\bibfnamefont {M.~M.}\ \bibnamefont {Becker}}, \bibinfo
  {author} {\bibfnamefont {T.}~\bibnamefont {Hoder}}, \bibinfo {author}
  {\bibfnamefont {D.}~\bibnamefont {Loffhagen}}, \bibinfo {author}
  {\bibfnamefont {R.}~\bibnamefont {Brandenburg}},\ and\ \bibinfo {author}
  {\bibfnamefont {K.-D.}\ \bibnamefont {Weltmann}},\ }\bibfield  {title}
  {\bibinfo {title} {Breakdown characteristics in pulsed-driven dielectric
  barrier discharges: influence of the pre-breakdown phase due to volume memory
  effects},\ }\href {https://doi.org/10.1088/0022-3727/47/46/465206} {\bibfield
   {journal} {\bibinfo  {journal} {Journal of Physics D: Applied Physics}\
  }\textbf {\bibinfo {volume} {47}},\ \bibinfo {pages} {465206} (\bibinfo
  {year} {2014})}\BibitemShut {NoStop}%
\bibitem [{\citenamefont {Nemschokmichal}\ \emph {et~al.}(2018)\citenamefont
  {Nemschokmichal}, \citenamefont {Tschiersch}, \citenamefont {Höft},
  \citenamefont {Wild}, \citenamefont {Bogaczyk}, \citenamefont {Becker},
  \citenamefont {Loffhagen}, \citenamefont {Stollenwerk}, \citenamefont
  {Kettlitz}, \citenamefont {Brandenburg},\ and\ \citenamefont
  {Meichsner}}]{Nemschokmichal-2018-ID5108}%
  \BibitemOpen
  \bibfield  {author} {\bibinfo {author} {\bibfnamefont {S.}~\bibnamefont
  {Nemschokmichal}}, \bibinfo {author} {\bibfnamefont {R.}~\bibnamefont
  {Tschiersch}}, \bibinfo {author} {\bibfnamefont {H.}~\bibnamefont {Höft}},
  \bibinfo {author} {\bibfnamefont {R.}~\bibnamefont {Wild}}, \bibinfo {author}
  {\bibfnamefont {M.}~\bibnamefont {Bogaczyk}}, \bibinfo {author}
  {\bibfnamefont {M.~M.}\ \bibnamefont {Becker}}, \bibinfo {author}
  {\bibfnamefont {D.}~\bibnamefont {Loffhagen}}, \bibinfo {author}
  {\bibfnamefont {L.}~\bibnamefont {Stollenwerk}}, \bibinfo {author}
  {\bibfnamefont {M.}~\bibnamefont {Kettlitz}}, \bibinfo {author}
  {\bibfnamefont {R.}~\bibnamefont {Brandenburg}},\ and\ \bibinfo {author}
  {\bibfnamefont {J.}~\bibnamefont {Meichsner}},\ }\bibfield  {title} {\bibinfo
  {title} {Impact of volume and surface processes on the pre-ionization of
  dielectric barrier discharges: advanced diagnostics and fluid modeling},\
  }\href {https://doi.org/10.1140/epjd/e2017-80369-1} {\bibfield  {journal}
  {\bibinfo  {journal} {The European Physical Journal D}\ }\textbf {\bibinfo
  {volume} {72}},\ \bibinfo {pages} {89} (\bibinfo {year} {2018})}\BibitemShut
  {NoStop}%
\bibitem [{\citenamefont {Kuthanová}\ and\ \citenamefont
  {Hoder}(2022)}]{kutha2022}%
  \BibitemOpen
  \bibfield  {author} {\bibinfo {author} {\bibfnamefont {L.}~\bibnamefont
  {Kuthanová}}\ and\ \bibinfo {author} {\bibfnamefont {T.}~\bibnamefont
  {Hoder}},\ }\bibfield  {title} {\bibinfo {title} {Memory propagation in
  barrier discharge at water interface: suspected markov states and
  spatiotemporal memory effects},\ }\href
  {https://doi.org/10.1088/1361-6595/ac64bb} {\bibfield  {journal} {\bibinfo
  {journal} {Plasma Sources Science and Technology}\ }\textbf {\bibinfo
  {volume} {31}},\ \bibinfo {pages} {045022} (\bibinfo {year}
  {2022})}\BibitemShut {NoStop}%
\bibitem [{\citenamefont {Li}\ \emph {et~al.}(2022)\citenamefont {Li},
  \citenamefont {Yan}, \citenamefont {Li}, \citenamefont {Schulze},
  \citenamefont {Yu}, \citenamefont {Song},\ and\ \citenamefont
  {Zhang}}]{li2022}%
  \BibitemOpen
  \bibfield  {author} {\bibinfo {author} {\bibfnamefont {T.}~\bibnamefont
  {Li}}, \bibinfo {author} {\bibfnamefont {H.-J.}\ \bibnamefont {Yan}},
  \bibinfo {author} {\bibfnamefont {J.-Q.}\ \bibnamefont {Li}}, \bibinfo
  {author} {\bibfnamefont {J.}~\bibnamefont {Schulze}}, \bibinfo {author}
  {\bibfnamefont {S.-Q.}\ \bibnamefont {Yu}}, \bibinfo {author} {\bibfnamefont
  {J.}~\bibnamefont {Song}},\ and\ \bibinfo {author} {\bibfnamefont {Q.-Z.}\
  \bibnamefont {Zhang}},\ }\bibfield  {title} {\bibinfo {title} {The role of
  surface charge and its decay in surface dielectric barrier discharges},\
  }\href {https://doi.org/10.1088/1361-6595/ac676e} {\bibfield  {journal}
  {\bibinfo  {journal} {Plasma Sources Science and Technology}\ }\textbf
  {\bibinfo {volume} {31}},\ \bibinfo {pages} {055016} (\bibinfo {year}
  {2022})}\BibitemShut {NoStop}%
\bibitem [{\citenamefont {Černák}\ \emph {et~al.}(2011)\citenamefont
  {Černák}, \citenamefont {Kováčik}, \citenamefont {Ráhel'}, \citenamefont
  {St'ahel}, \citenamefont {Zahoranová}, \citenamefont {Kubincová},
  \citenamefont {Tóth},\ and\ \citenamefont {Černáková}}]{cernak2011}%
  \BibitemOpen
  \bibfield  {author} {\bibinfo {author} {\bibfnamefont {M.}~\bibnamefont
  {Černák}}, \bibinfo {author} {\bibfnamefont {D.}~\bibnamefont {Kováčik}},
  \bibinfo {author} {\bibfnamefont {J.}~\bibnamefont {Ráhel'}}, \bibinfo
  {author} {\bibfnamefont {P.}~\bibnamefont {St'ahel}}, \bibinfo {author}
  {\bibfnamefont {A.}~\bibnamefont {Zahoranová}}, \bibinfo {author}
  {\bibfnamefont {J.}~\bibnamefont {Kubincová}}, \bibinfo {author}
  {\bibfnamefont {A.}~\bibnamefont {Tóth}},\ and\ \bibinfo {author}
  {\bibfnamefont {L.}~\bibnamefont {Černáková}},\ }\bibfield  {title}
  {\bibinfo {title} {Generation of a high-density highly non-equilibrium air
  plasma for high-speed large-area flat surface processing},\ }\href
  {https://doi.org/10.1088/0741-3335/53/12/124031} {\bibfield  {journal}
  {\bibinfo  {journal} {Plasma Physics and Controlled Fusion}\ }\textbf
  {\bibinfo {volume} {53}},\ \bibinfo {pages} {124031} (\bibinfo {year}
  {2011})}\BibitemShut {NoStop}%
\bibitem [{\citenamefont {Douat}\ \emph {et~al.}(2023)\citenamefont {Douat},
  \citenamefont {Ponduri}, \citenamefont {Boumans}, \citenamefont {Guaitella},
  \citenamefont {Welzel}, \citenamefont {Carbone},\ and\ \citenamefont
  {Engeln}}]{douat2023}%
  \BibitemOpen
  \bibfield  {author} {\bibinfo {author} {\bibfnamefont {C.}~\bibnamefont
  {Douat}}, \bibinfo {author} {\bibfnamefont {S.}~\bibnamefont {Ponduri}},
  \bibinfo {author} {\bibfnamefont {T.}~\bibnamefont {Boumans}}, \bibinfo
  {author} {\bibfnamefont {O.}~\bibnamefont {Guaitella}}, \bibinfo {author}
  {\bibfnamefont {S.}~\bibnamefont {Welzel}}, \bibinfo {author} {\bibfnamefont
  {E.}~\bibnamefont {Carbone}},\ and\ \bibinfo {author} {\bibfnamefont
  {R.}~\bibnamefont {Engeln}},\ }\bibfield  {title} {\bibinfo {title} {The role
  of the number of filaments in the dissociation of {CO}\textsubscript{2} in
  dielectric barrier discharges},\ }\href
  {https://doi.org/10.1088/1361-6595/acceca} {\bibfield  {journal} {\bibinfo
  {journal} {Plasma Sources Science and Technology}\ }\textbf {\bibinfo
  {volume} {32}},\ \bibinfo {pages} {055001} (\bibinfo {year}
  {2023})}\BibitemShut {NoStop}%
\bibitem [{\citenamefont {Liu}\ and\ \citenamefont
  {Neiger}(2003)}]{liu_electrical_2003}%
  \BibitemOpen
  \bibfield  {author} {\bibinfo {author} {\bibfnamefont {S.}~\bibnamefont
  {Liu}}\ and\ \bibinfo {author} {\bibfnamefont {M.}~\bibnamefont {Neiger}},\
  }\bibfield  {title} {\bibinfo {title} {Electrical modelling of homogeneous
  dielectric barrier discharges under an arbitrary excitation voltage},\ }\href
  {https://doi.org/10.1088/0022-3727/36/24/009} {\bibfield  {journal} {\bibinfo
   {journal} {Journal of Physics D: Applied Physics}\ }\textbf {\bibinfo
  {volume} {36}},\ \bibinfo {pages} {3144} (\bibinfo {year}
  {2003})}\BibitemShut {NoStop}%
\bibitem [{\citenamefont {Pipa}\ \emph {et~al.}(2012)\citenamefont {Pipa},
  \citenamefont {Koskulics}, \citenamefont {Brandenburg},\ and\ \citenamefont
  {Hoder}}]{pipa_simplest_2012}%
  \BibitemOpen
  \bibfield  {author} {\bibinfo {author} {\bibfnamefont {A.~V.}\ \bibnamefont
  {Pipa}}, \bibinfo {author} {\bibfnamefont {J.}~\bibnamefont {Koskulics}},
  \bibinfo {author} {\bibfnamefont {R.}~\bibnamefont {Brandenburg}},\ and\
  \bibinfo {author} {\bibfnamefont {T.}~\bibnamefont {Hoder}},\ }\bibfield
  {title} {\bibinfo {title} {The simplest equivalent circuit of a pulsed
  dielectric barrier discharge and the determination of the gas gap charge
  transfer},\ }\href {https://doi.org/10.1063/1.4767637} {\bibfield  {journal}
  {\bibinfo  {journal} {Review of Scientific Instruments}\ }\textbf {\bibinfo
  {volume} {83}},\ \bibinfo {pages} {115112} (\bibinfo {year}
  {2012})}\BibitemShut {NoStop}%
\bibitem [{\citenamefont {Ivković}\ \emph {et~al.}(2009)\citenamefont
  {Ivković}, \citenamefont {Obradović}, \citenamefont {Cvetanović},
  \citenamefont {Kuraica},\ and\ \citenamefont {Purić}}]{ivkovic2009}%
  \BibitemOpen
  \bibfield  {author} {\bibinfo {author} {\bibfnamefont {S.~S.}\ \bibnamefont
  {Ivković}}, \bibinfo {author} {\bibfnamefont {B.~M.}\ \bibnamefont
  {Obradović}}, \bibinfo {author} {\bibfnamefont {N.}~\bibnamefont
  {Cvetanović}}, \bibinfo {author} {\bibfnamefont {M.~M.}\ \bibnamefont
  {Kuraica}},\ and\ \bibinfo {author} {\bibfnamefont {J.}~\bibnamefont
  {Purić}},\ }\bibfield  {title} {\bibinfo {title} {Measurement of electric
  field development in dielectric barrier discharge in helium},\ }\href
  {https://doi.org/10.1088/0022-3727/42/22/225206} {\bibfield  {journal}
  {\bibinfo  {journal} {Journal of Physics D: Applied Physics}\ }\textbf
  {\bibinfo {volume} {42}},\ \bibinfo {pages} {225206} (\bibinfo {year}
  {2009})}\BibitemShut {NoStop}%
\bibitem [{\citenamefont {Mrkvičková}\ \emph {et~al.}(2023)\citenamefont
  {Mrkvičková}, \citenamefont {Kuthanová}, \citenamefont {Bílek},
  \citenamefont {Obrusník}, \citenamefont {Navrátil}, \citenamefont
  {Dvořák}, \citenamefont {Adamovich}, \citenamefont {Šimek},\ and\
  \citenamefont {Hoder}}]{mrkvickova2023}%
  \BibitemOpen
  \bibfield  {author} {\bibinfo {author} {\bibfnamefont {M.}~\bibnamefont
  {Mrkvičková}}, \bibinfo {author} {\bibfnamefont {L.}~\bibnamefont
  {Kuthanová}}, \bibinfo {author} {\bibfnamefont {P.}~\bibnamefont {Bílek}},
  \bibinfo {author} {\bibfnamefont {A.}~\bibnamefont {Obrusník}}, \bibinfo
  {author} {\bibfnamefont {Z.}~\bibnamefont {Navrátil}}, \bibinfo {author}
  {\bibfnamefont {P.}~\bibnamefont {Dvořák}}, \bibinfo {author}
  {\bibfnamefont {I.}~\bibnamefont {Adamovich}}, \bibinfo {author}
  {\bibfnamefont {M.}~\bibnamefont {Šimek}},\ and\ \bibinfo {author}
  {\bibfnamefont {T.}~\bibnamefont {Hoder}},\ }\bibfield  {title} {\bibinfo
  {title} {Electric field in {APTD} in nitrogen determined by {EFISH},
  {FNS}/{SPS} ratio, $\alpha$-fitting and electrical equivalent circuit
  model},\ }\href {https://doi.org/10.1088/1361-6595/acd6de} {\bibfield
  {journal} {\bibinfo  {journal} {Plasma Sources Science and Technology}\
  }\textbf {\bibinfo {volume} {32}},\ \bibinfo {pages} {065009} (\bibinfo
  {year} {2023})}\BibitemShut {NoStop}%
\bibitem [{\citenamefont {Peeters}\ and\ \citenamefont {van~de
  Sanden}(2014)}]{peeters2015}%
  \BibitemOpen
  \bibfield  {author} {\bibinfo {author} {\bibfnamefont {F.~J.~J.}\
  \bibnamefont {Peeters}}\ and\ \bibinfo {author} {\bibfnamefont {M.~C.~M.}\
  \bibnamefont {van~de Sanden}},\ }\bibfield  {title} {\bibinfo {title} {The
  influence of partial surface discharging on the electrical characterization
  of dbds},\ }\href {https://doi.org/10.1088/0963-0252/24/1/015016} {\bibfield
  {journal} {\bibinfo  {journal} {Plasma Sources Science and Technology}\
  }\textbf {\bibinfo {volume} {24}},\ \bibinfo {pages} {015016} (\bibinfo
  {year} {2014})}\BibitemShut {NoStop}%
\bibitem [{\citenamefont {Akishev}\ \emph {et~al.}(2011)\citenamefont
  {Akishev}, \citenamefont {Aponin}, \citenamefont {Balakirev}, \citenamefont
  {Grushin}, \citenamefont {Karalnik}, \citenamefont {Petryakov},\ and\
  \citenamefont {Trushkin}}]{akishev_memory_2011}%
  \BibitemOpen
  \bibfield  {author} {\bibinfo {author} {\bibfnamefont {Y.}~\bibnamefont
  {Akishev}}, \bibinfo {author} {\bibfnamefont {G.}~\bibnamefont {Aponin}},
  \bibinfo {author} {\bibfnamefont {A.}~\bibnamefont {Balakirev}}, \bibinfo
  {author} {\bibfnamefont {M.}~\bibnamefont {Grushin}}, \bibinfo {author}
  {\bibfnamefont {V.}~\bibnamefont {Karalnik}}, \bibinfo {author}
  {\bibfnamefont {A.}~\bibnamefont {Petryakov}},\ and\ \bibinfo {author}
  {\bibfnamefont {N.}~\bibnamefont {Trushkin}},\ }\bibfield  {title} {\bibinfo
  {title} {‘{Memory}’ and sustention of microdischarges in a steady-state
  {DBD}: volume plasma or surface charge?},\ }\href
  {https://doi.org/10.1088/0963-0252/20/2/024005} {\bibfield  {journal}
  {\bibinfo  {journal} {Plasma Sources Science and Technology}\ }\textbf
  {\bibinfo {volume} {20}},\ \bibinfo {pages} {024005} (\bibinfo {year}
  {2011})}\BibitemShut {NoStop}%
\bibitem [{\citenamefont {Tyl}\ \emph {et~al.}(2021)\citenamefont {Tyl},
  \citenamefont {Martin}, \citenamefont {Combettes}, \citenamefont {Brillat},
  \citenamefont {Bley}, \citenamefont {Belinger}, \citenamefont {Dap},
  \citenamefont {Brandenburg},\ and\ \citenamefont {Naudé}}]{tyl2021}%
  \BibitemOpen
  \bibfield  {author} {\bibinfo {author} {\bibfnamefont {C.}~\bibnamefont
  {Tyl}}, \bibinfo {author} {\bibfnamefont {S.}~\bibnamefont {Martin}},
  \bibinfo {author} {\bibfnamefont {C.}~\bibnamefont {Combettes}}, \bibinfo
  {author} {\bibfnamefont {G.}~\bibnamefont {Brillat}}, \bibinfo {author}
  {\bibfnamefont {V.}~\bibnamefont {Bley}}, \bibinfo {author} {\bibfnamefont
  {A.}~\bibnamefont {Belinger}}, \bibinfo {author} {\bibfnamefont
  {S.}~\bibnamefont {Dap}}, \bibinfo {author} {\bibfnamefont {R.}~\bibnamefont
  {Brandenburg}},\ and\ \bibinfo {author} {\bibfnamefont {N.}~\bibnamefont
  {Naudé}},\ }\bibfield  {title} {\bibinfo {title} {New local electrical
  diagnostic tool for dielectric barrier discharge ({DBD})},\ }\href
  {https://doi.org/10.1063/5.0045654} {\bibfield  {journal} {\bibinfo
  {journal} {Review of Scientific Instruments}\ }\textbf {\bibinfo {volume}
  {92}},\ \bibinfo {pages} {053552} (\bibinfo {year} {2021})}\BibitemShut
  {NoStop}%
\bibitem [{\citenamefont {Willebrand}\ \emph {et~al.}(1991)\citenamefont
  {Willebrand}, \citenamefont {Niedernostheide}, \citenamefont {Ammelt},
  \citenamefont {Dohmen},\ and\ \citenamefont
  {Purwins}}]{willebrand_spatio-temporal_1991}%
  \BibitemOpen
  \bibfield  {author} {\bibinfo {author} {\bibfnamefont {H.}~\bibnamefont
  {Willebrand}}, \bibinfo {author} {\bibfnamefont {F.-J.}\ \bibnamefont
  {Niedernostheide}}, \bibinfo {author} {\bibfnamefont {E.}~\bibnamefont
  {Ammelt}}, \bibinfo {author} {\bibfnamefont {R.}~\bibnamefont {Dohmen}},\
  and\ \bibinfo {author} {\bibfnamefont {H.-G.}\ \bibnamefont {Purwins}},\
  }\bibfield  {title} {\bibinfo {title} {Spatio-temporal oscillations during
  filament splitting in gas discharge systems},\ }\href
  {https://doi.org/https://doi.org/10.1016/0375-9601(91)90693-3} {\bibfield
  {journal} {\bibinfo  {journal} {Physics Letters A}\ }\textbf {\bibinfo
  {volume} {153}},\ \bibinfo {pages} {437} (\bibinfo {year}
  {1991})}\BibitemShut {NoStop}%
\bibitem [{\citenamefont {Boeuf}\ \emph {et~al.}(2012)\citenamefont {Boeuf},
  \citenamefont {Bernecker}, \citenamefont {Callegari}, \citenamefont
  {Blanco},\ and\ \citenamefont {Fournier}}]{boeuf_generation_2012}%
  \BibitemOpen
  \bibfield  {author} {\bibinfo {author} {\bibfnamefont {J.~P.}\ \bibnamefont
  {Boeuf}}, \bibinfo {author} {\bibfnamefont {B.}~\bibnamefont {Bernecker}},
  \bibinfo {author} {\bibfnamefont {T.}~\bibnamefont {Callegari}}, \bibinfo
  {author} {\bibfnamefont {S.}~\bibnamefont {Blanco}},\ and\ \bibinfo {author}
  {\bibfnamefont {R.}~\bibnamefont {Fournier}},\ }\bibfield  {title} {\bibinfo
  {title} {Generation, annihilation, dynamics and self-organized patterns of
  filaments in dielectric barrier discharge plasmas},\ }\href
  {https://doi.org/10.1063/1.4729767} {\bibfield  {journal} {\bibinfo
  {journal} {Applied Physics Letters}\ }\textbf {\bibinfo {volume} {100}},\
  \bibinfo {pages} {244108} (\bibinfo {year} {2012})}\BibitemShut {NoStop}%
\bibitem [{\citenamefont {Pipa}\ and\ \citenamefont
  {Brandenburg}(2019)}]{pipa2019}%
  \BibitemOpen
  \bibfield  {author} {\bibinfo {author} {\bibfnamefont {A.~V.}\ \bibnamefont
  {Pipa}}\ and\ \bibinfo {author} {\bibfnamefont {R.}~\bibnamefont
  {Brandenburg}},\ }\bibfield  {title} {\bibinfo {title} {The equivalent
  circuit approach for the electrical diagnostics of dielectric barrier
  discharges: The classical theory and recent developments},\ }\href
  {https://doi.org/10.3390/atoms7010014} {\bibfield  {journal} {\bibinfo
  {journal} {Atoms}\ }\textbf {\bibinfo {volume} {7}},\ \bibinfo {pages} {14}
  (\bibinfo {year} {2019})}\BibitemShut {NoStop}%
\bibitem [{\citenamefont {Sewraj}\ \emph {et~al.}(2011)\citenamefont {Sewraj},
  \citenamefont {Merbahi}, \citenamefont {Gardou}, \citenamefont {Akerreta},\
  and\ \citenamefont {Marchal}}]{sewraj2011}%
  \BibitemOpen
  \bibfield  {author} {\bibinfo {author} {\bibfnamefont {N.}~\bibnamefont
  {Sewraj}}, \bibinfo {author} {\bibfnamefont {N.}~\bibnamefont {Merbahi}},
  \bibinfo {author} {\bibfnamefont {J.~P.}\ \bibnamefont {Gardou}}, \bibinfo
  {author} {\bibfnamefont {P.~R.}\ \bibnamefont {Akerreta}},\ and\ \bibinfo
  {author} {\bibfnamefont {F.}~\bibnamefont {Marchal}},\ }\bibfield  {title}
  {\bibinfo {title} {Electric and spectroscopic analysis of a pure nitrogen
  mono-filamentary dielectric barrier discharge ({MF}-{DBD}) at 760 {Torr}},\
  }\href {https://doi.org/10.1088/0022-3727/44/14/145201} {\bibfield  {journal}
  {\bibinfo  {journal} {Journal of Physics D: Applied Physics}\ }\textbf
  {\bibinfo {volume} {44}},\ \bibinfo {pages} {145201} (\bibinfo {year}
  {2011})}\BibitemShut {NoStop}%
\bibitem [{\citenamefont {Stankov}\ \emph {et~al.}(2026)\citenamefont
  {Stankov}, \citenamefont {Boer}, \citenamefont {Graef}, \citenamefont {{van
  ’t Veer}}, \citenamefont {Jovanović}, \citenamefont {Sigeneger},
  \citenamefont {Loffhagen}, \citenamefont {{van Dijk}},\ and\ \citenamefont
  {Becker}}]{stankov_mcplas_2026}%
  \BibitemOpen
  \bibfield  {author} {\bibinfo {author} {\bibfnamefont {M.}~\bibnamefont
  {Stankov}}, \bibinfo {author} {\bibfnamefont {D.}~\bibnamefont {Boer}},
  \bibinfo {author} {\bibfnamefont {W.}~\bibnamefont {Graef}}, \bibinfo
  {author} {\bibfnamefont {K.}~\bibnamefont {{van ’t Veer}}}, \bibinfo
  {author} {\bibfnamefont {A.~P.}\ \bibnamefont {Jovanović}}, \bibinfo
  {author} {\bibfnamefont {F.}~\bibnamefont {Sigeneger}}, \bibinfo {author}
  {\bibfnamefont {D.}~\bibnamefont {Loffhagen}}, \bibinfo {author}
  {\bibfnamefont {J.}~\bibnamefont {{van Dijk}}},\ and\ \bibinfo {author}
  {\bibfnamefont {M.~M.}\ \bibnamefont {Becker}},\ }\bibfield  {title}
  {\bibinfo {title} {{MCPlas}, a {MATLAB} toolbox for reproducible plasma
  modelling with {COMSOL}},\ }\href
  {https://doi.org/https://doi.org/10.1016/j.cpc.2026.110248} {\bibfield
  {journal} {\bibinfo  {journal} {Computer Physics Communications}\ }\textbf
  {\bibinfo {volume} {327}},\ \bibinfo {pages} {110248} (\bibinfo {year}
  {2026})}\BibitemShut {NoStop}%
\bibitem [{\citenamefont {Höft}\ \emph {et~al.}(2022)\citenamefont {Höft},
  \citenamefont {Becker}, \citenamefont {Kettlitz},\ and\ \citenamefont
  {Brandenburg}}]{hoft_upscaling_2022}%
  \BibitemOpen
  \bibfield  {author} {\bibinfo {author} {\bibfnamefont {H.}~\bibnamefont
  {Höft}}, \bibinfo {author} {\bibfnamefont {M.~M.}\ \bibnamefont {Becker}},
  \bibinfo {author} {\bibfnamefont {M.}~\bibnamefont {Kettlitz}},\ and\
  \bibinfo {author} {\bibfnamefont {R.}~\bibnamefont {Brandenburg}},\
  }\bibfield  {title} {\bibinfo {title} {Upscaling from single- to
  multi-filament dielectric barrier discharges in pulsed operation},\ }\href
  {https://doi.org/10.1088/1361-6463/ac868b} {\bibfield  {journal} {\bibinfo
  {journal} {Journal of Physics D: Applied Physics}\ }\textbf {\bibinfo
  {volume} {55}},\ \bibinfo {pages} {424003} (\bibinfo {year}
  {2022})}\BibitemShut {NoStop}%
\end{thebibliography}%
